\documentclass[conference]{IEEEtran}
\IEEEoverridecommandlockouts
\usepackage{cite}
\usepackage{amsmath,amssymb,amsfonts}
\usepackage{algorithmic}
\usepackage{graphicx}
\usepackage{subcaption}
\usepackage{textcomp}
\usepackage{xcolor}
\usepackage{booktabs}
\usepackage[skins]{tcolorbox}
\usepackage{hyperref}
\usepackage{multirow}
\usepackage{enumitem}
\usepackage{balance}
\usepackage{bbding}
\usepackage{utfsym}
\usepackage{ulem}
\usepackage{hyperref}
\usepackage[linesnumbered,ruled,vlined]{algorithm2e}
\usepackage{subcaption}
\usepackage{makecell}
\usepackage{cleveref}

\def\BibTeX{{\rm B\kern-.05em{\sc i\kern-.025em b}\kern-.08em
    T\kern-.1667em\lower.7ex\hbox{E}\kern-.125emX}}

\tcbset{
  my box/.style={
    enhanced,
    colframe=#1!80,
    colback=#1!10,
    attach boxed title to top left={xshift=0.2cm, yshift=-0.2cm},
    boxed title style={
      colback=#1!80,
      outer arc=0pt,
      arc=0pt,
      top=0pt,
      bottom=0pt,
    },
  },
}
\newtcolorbox{RQResult}[1]{
  my box=black,
  title=#1,
  boxrule=1.2pt,top=6pt,bottom=3.5pt,left=6pt,right=6pt
}

\makeatletter
\newcommand{\linebreakand}{%
  \end{@IEEEauthorhalign}
  \hfill\mbox{}\par
  \mbox{}\hfill\begin{@IEEEauthorhalign}
}
\makeatother

\begin{document}

\title{Decomposing God Header File via Multi-View Graph Clustering
}


\author{
\IEEEauthorblockN{Yue Wang}
\IEEEauthorblockA{\textit{Key Lab of HCST (PKU), MOE;}\\
\textit{SCS, Peking University}\\
Beijing, China \\
wangyue0502@pku.edu.cn}
\and
\IEEEauthorblockN{Wenhui Chang}
\IEEEauthorblockA{\textit{Key Lab of HCST (PKU), MOE;}\\
\textit{SCS, Peking University}\\
Beijing, China \\
wenhui\_chang@pku.edu.cn}
\and
\IEEEauthorblockN{Tongwei Deng}
\IEEEauthorblockA{\textit{Key Lab of HCST (PKU), MOE;}\\
\textit{SCS, Peking University}\\
Beijing, China \\
dtw@stu.pku.edu.cn}
\linebreakand 
\IEEEauthorblockN{Yanzhen Zou*\thanks{* Corresponding author.}}
\IEEEauthorblockA{\textit{Key Lab of HCST (PKU), MOE;}\\
\textit{SCS, Peking University}\\
Beijing, China \\
zouyz@pku.edu.cn}
\and
\IEEEauthorblockN{Bing Xie}
\IEEEauthorblockA{\textit{Key Lab of HCST (PKU), MOE;}\\
\textit{SCS, Peking University}\\
Beijing, China \\
xiebing@pku.edu.cn}
}

\maketitle

\begin{abstract}
God Header Files, just like God Classes, pose significant challenges for code comprehension and maintenance. Additionally, they increase the time required for code recompilation. However, existing refactoring methods for God Classes are inappropriate to deal with God Header Files because the code elements in header files are mostly short declaration types, and build dependencies of the entire system should be considered with the aim of improving compilation efficiency. Meanwhile, ensuring acyclic dependencies among the decomposed sub-header files is also crucial in the God Header File decomposition. 
This paper proposes a multi-view graph clustering based approach for decomposing God Header Files. It first constructs and coarsens the code element graph, then a novel multi-view graph clustering algorithm is applied to identify the clusters and a heuristic algorithm is introduced to address the cyclic dependencies in the clustering results. 
To evaluate our approach, we built both a synthetic dataset and a real-world God Header Files dataset. The results show that 1) Our approach could achieve 11.5\% higher accuracy than existing God Class refactoring methods; 2) Our decomposition results attain better architecture on real-world God Header Files, evidenced by higher modularity and acyclic dependencies; 3) We can reduce 15\% to 60\% recompilation time for historical commits that require recompiling.
\end{abstract}


\section{Introduction}
\label{sec:introduction}


Code refactoring plays a crucial role in long-lifespan software projects. 
As the software is enhanced, modified, and adapted to new requirements, the code becomes more complex and drifts away from its original design, thereby degrading the quality of the software~\cite{mens2004survey}.
Refactoring enables developers to transform poorly designed or chaotic code into well-structured one~\cite{fowler2018refactoring}, thus improving the software's robustness, reusability, performance, and other essential attributes.

\begin{figure}[!t]
  \centering
  \includegraphics[width=\linewidth]{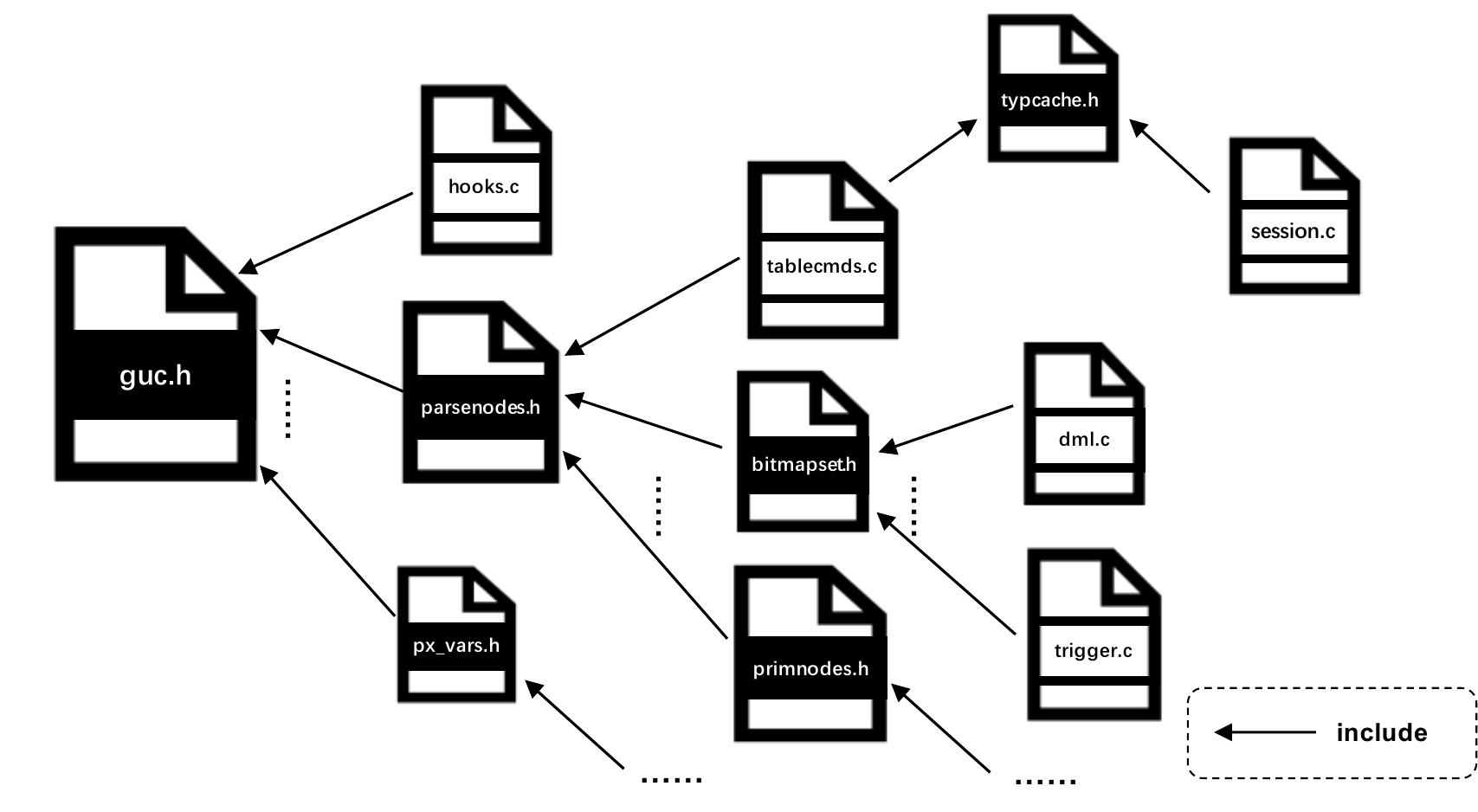}
  \caption{An example of God Header File (\texttt{guc.h}).}
  \label{fig:godheaderfileexample}
  \vspace{-5pt}
\end{figure}

During our collaboration with an embedded software development enterprise, we encounter a pressing code refactoring problem, named God Header File, where some header files exhibit large code size and wide file impact in the C language projects. Once such a header file is modified in a commit, all files that include it must be recompiled too. Eventually, the recompilation time of the related code commit becomes unbearable.  
Figure~\ref{fig:godheaderfileexample} shows an example of such header files. The file \texttt{guc.h} is a header file from a cloud-native database software project PolarDB-for-PostgreSQL~\cite{PolarDB-for-PostgreSQL}. It contains 63 macro definitions, 19 data structure definitions, and 493 declarations of variables or functions. Any change to it triggers recompilation of 388 files directly including it (such as \texttt{hooks.c}) or transitively including it through other header files (such as \texttt{tablecmds.c} and \texttt{dml.c}), amounting to a total of 741,265 lines of code, incurring significant recompilation cost. 
Therefore, it's necessary to refactor these God Header Files so as to optimize software structure and reduce compilation cost during software evolution.

To deal with God Header Files, we try to draw inspiration from the well-recognized tasks of God Class refactoring. God Class~\cite{riel1996object} refers to a large class with many responsibilities in a system, which poses difficulties for developers in code comprehension, test, and maintenance~\cite{alzahrani2022extract}. To deal with God Classes, some approaches have been proposed to refactor them by extracting new classes automatically, including static analysis-based methods~\cite{bavota2010playing, bavota2011identifying, bavota2014automating}, metric-based methods~\cite{jeba2020god, akash2019approach} and deep learning-based methods~\cite{akash2022exploring}. These approaches pay attention to optimizing the software structure, resulting in refactored classes with higher cohesion and lower coupling.

However, we found that these God Class refactoring methods are not suitable for decomposing God Header Files due to the following three reasons. 
First, existing approaches primarily focus on the internal dependencies among methods in a God Class, but we have to take into account build dependencies~\cite{mcintosh2016identifying} that lie outside the God Header File in order to improve compilation efficiency. 
Second, existing methods usually combine multiple code relationships through weighted summation. However, determining the weights for each type of relationship is quite challenging and the weights might vary across different projects~\cite{bavota2014automating}. 
Last, these methods have overlooked the concern of cyclic dependencies among decomposed files, which is a kind of architectural anti-pattern~\cite{taibi2018definition} and particularly intolerant in header files as they may lead to compilation errors.

To address the above challenges, we propose a God Header File decomposing method based on multi-view clustering that leverages different types of code relationships. 
First, we construct a code element graph considering not only the internal dependency and semantic relationships among code elements in a God Header File, but also the co-usage relationships, which reflect build dependencies of the entire project. 
Then, we coarsen the graph based on dependency relationships and employ a multi-view graph clustering algorithm to cluster code elements based on their semantic and co-usage relationships. 
Finally, a heuristic algorithm is introduced to address the cyclic dependencies in the clustering result.

We evaluate our approach on both synthetic God Header Files and real-world God Header Files. The real-world God Header Files are derived from a preliminary study of 557 projects on GitHub. The experimental results reveal that:
1) our approach achieves 11.5\% higher accuracy in comparison to existing methods, and exhibits more consistent performance across different projects; 
2) when applied to real-world God Header Files, our approach provides decomposition results with better modularity and acyclic dependencies; 
3) our approach can reduce the recompilation cost for historical commits by 15\% to 60\%.



Compared with existing work, this paper gives the following contributions:
\begin{itemize} 
    \item   Make aware of the problem of God Header Files. We carried out a study on the C language projects in GitHub, which shows that God Header Files are widespread in open-source projects.
    
    \item   Propose a God Header File decomposing approach via multi-view graph clustering, which achieves better accuracy and modularity than existing methods. Meanwhile, it can reduce 15\% to 60\% recompilation time for the related historical commits.
    
    
    \item  Build a benchmark to investigate the effectiveness of God Header File decomposition approaches, which is publicly available~\cite{web:dataset} to facilitate future work in this area.
\end{itemize}
\section{Preliminary Study on God Header Files}
\label{subsec:preliminary}
Here we conduct a preliminary study to investigate the prevalence of God Header Files in open-source projects. 

\subsection{Study Setup}

To carry out the study, we collected the open-source C language projects and analyzed their header files. We manually annotated some samples to estimate the spread of God Header Files on GitHub. The study includes the following steps:

\textbf{Projects Selection}. We collected C language projects with size of more than 10,000 KB and stars of more than 500 from GitHub. Then we obtained 557 projects (downloaded by January 2023).

\begin{figure}[!t]
  \centering
  \includegraphics[width=\linewidth]{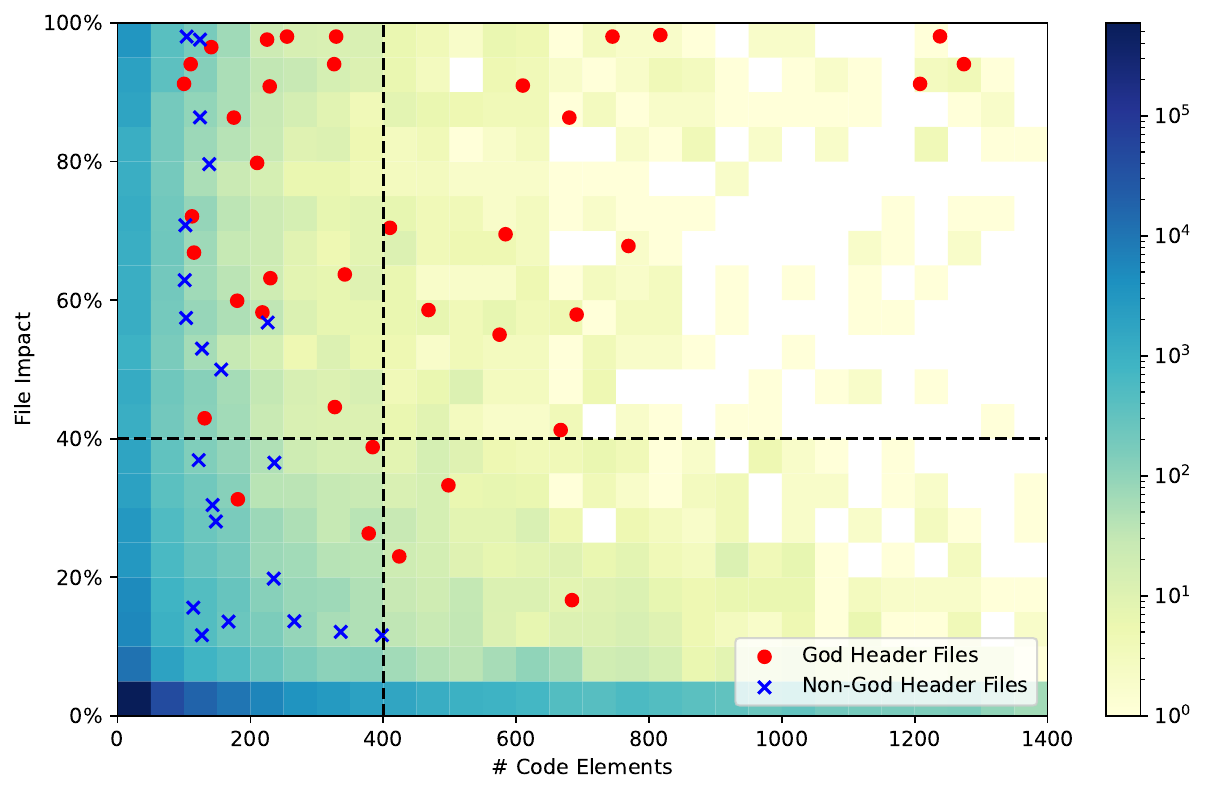}
  \caption{Joint distribution of header files' code size and file impact.}
  \label{fig:joint_distribution}
  \vspace{-5pt}
\end{figure}

\textbf{Header Files Analysis}. We used tree-sitter~\cite{tree-sitter}, a parser generator tool, to analyze header files in collected projects. For each header file, we calculated its code size and file impact. Code size is represented by the number of code elements (definitions/declarations of macros, data types, variables, and functions) in a header file. And file impact is calculated by the percentage of code lines in all impacted files to the total code lines in the project, where impacted files refer to the files that include the header file directly or transitively. 
Tree-sitter successfully parsed 541 projects, which comprised a total of 761,999 header files. And the joint distribution of code size and file impact of header files is shown in Figure~\ref{fig:joint_distribution}. The plot reveals that only a very small proportion of header files have both large code size and wide file impact.

\begin{table*}[t]
\centering
\caption{Projects for manual annotation and their typical God Header Files.}
\label{projects_and_headerfiles}
\resizebox{\linewidth}{!}{
\begin{tabular}{l|l|r|r|l|r|r|r}
    \toprule
    Project & Project Domain & Size(KB) &Stars &\makecell[l]{Typical God Header Files}   & \makecell[l]{\#Code \\elements}   & \makecell[l]{File \\Impact(\%)}  & \#Commits \\
    \midrule
    FreeRDP~\cite{FreeRDP} & remote desktop protocol &57,040 &7742 &settings.h & 745 & 98.0 & 309 \\
    \makecell[l]{PolarDB-for-PostgreSQL~\cite{PolarDB-for-PostgreSQL}} & cloud-native database &389,425 &2499 &guc.h & 575 & 55.0 & 61 \\
    SDL~\cite{SDL} & cross-platform multi-media library &140,832 &584 &\makecell[l]{SDL\_dynapi\_overrides.h} & 769 &67.8 & 112 \\
    SoftEtherVPN~\cite{VPN} & cross-platform multi-protocol VPN &540,135 &9758 &Network.h & 680  & 86.3 & 50 \\
    stress-ng~\cite{stress-ng} & system stress testing tool &25,993 &838 &stress-ng.h & 610  & 90.9 & 350  \\
    wiredtiger~\cite{wiredtiger} & data management platform &126,937 &1974 &extern.h & 1274 & 94.0 & 1364 \\
    \bottomrule
\end{tabular}
}
\vspace{-5pt}
\end{table*}

\textbf{Header Files Sampling and Annotation}. We selected 6 well-documented projects for manual annotation (shown in Table~\ref{projects_and_headerfiles}). We sampled header files from each project using a stratified approach:
for each range defined by $i*100 <$ code size $< (i+1)*100$ AND $j*0.1*100\% <$ file impact $< (j+1)*0.1*100\%$, where $i>1$ and $1<j<10$, we randomly selected one header file if available. 
This ensured a balanced sampling of files across varying ranges of code size and file impact.
And a total of 59 samples were collected in this process.
After that, two graduate students independently classified each sample as either a God Header File or not. In cases of disagreement, final adjudication was undertaken by the authors. 
The annotation results are depicted as scatter points in Figure~\ref{fig:joint_distribution}. The red circles represent God Header Files and the blue crosses are Non-God Header Files. It indicates that header files with larger code size wider file impact are more likely to be God Header Files.

\textbf{God Header Files Mining}. Based on the above findings, we regard header files with more than 400 code elements AND file impact exceeding 40\% of the its project as God Header Files in our subsequent experiment. 
The thresholds we picked (dotted lines in Figure~\ref{fig:joint_distribution}) are relatively high to ensure certainty in the identification of God Header Files, although may overlook some real ones.

\subsection{Study Results}

Based on the above work, we identified 649 God Header Files from 203 projects. 
These files constitute less than 1\% of the 761,999 total header files, which could be attributed to the strict thresholds we selected.
As for projects, 203 out of the 541 software projects contain God Header Files, indicating that at least 37.5\% of the projects in our study suffer from this issue. 
Specifically, 163 projects have 1 to 3 God Header Files. There are also 4 projects containing more than 20 God Header Files. Our further investigation revealed that these projects created duplicate header files to support multiple versions of hardware devices, contributing to the higher counts.

We also examined the commit histories of the 649 identified God Header Files. Among them, 103 files had more than 100 commits, and 14 files had been modified more than 5,000 times. Their large code size, wide file impact and frequent modifications increase the burden for software maintenance and evolution, underscoring the pressing need for a dedicated God Header File refactoring method.

In summary, we can conclude that \textbf{God Header Files are widespread in open-source C language projects}. Many of these files have undergone frequent modifications throughout their lifespan, incurring significant recompilation overhead. This highlights the need for researchers and developers to pay more attention to God Header File refactoring in the future.
Based on the study, we also built a dataset of God Header Files, including all the data used in this preliminary study. It is available at~\cite{web:dataset}.





\section{Approach}
\label{sec:approach}

\begin{figure*}[!t]
    \centering
    \includegraphics[width=\linewidth]{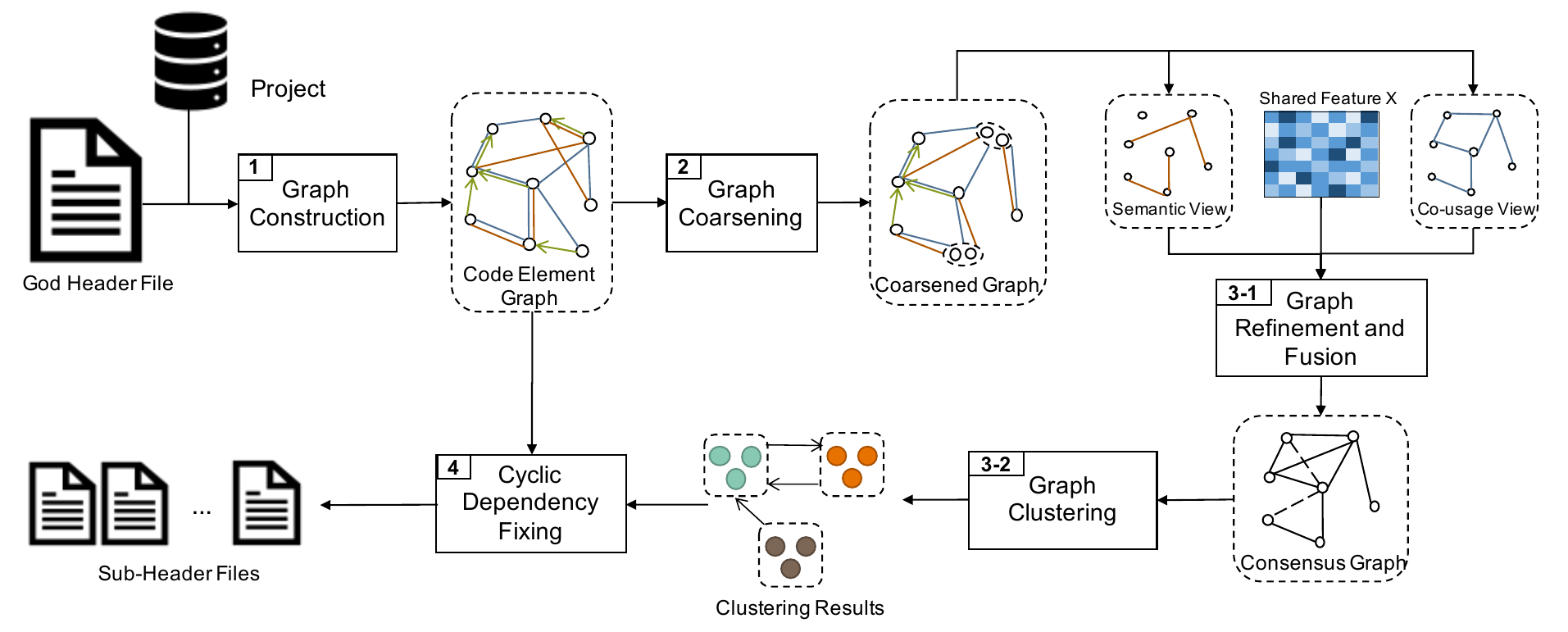}
    \caption{Overview of our God Header File Decomposition Approach.}
    \label{fig:approach_overview}
    \vspace{-5pt}
\end{figure*}

Figure~\ref{fig:approach_overview} presents an overview of the proposed approach that aims to automatically decompose a God Header File into several sub-header files. It can be divided into four parts: graph construction, graph coarsening, multi-view graph clustering, and cyclic dependency fixing.
In the \textbf{graph construction} phase (\cref{sec:stage1}), we parse the God Header File and extract code elements along with their relationships to construct the code element graph. The code elements are macros, data types, and function declarations, which cannot be further decomposed. The code relationships include dependency, semantic, and co-usage.
Then in the \textbf{graph coarsening} phase (\cref{sec:stage2}), we use the scarce but influential dependency relationships to coarsen the code element graph, ensuring that closely related code elements will not be separated in subsequent processes.
In the \textbf{multi-view clustering} part (\cref{sec:stage3}), we fuse the information from semantic and co-usage relationships, which exhibits more intricate but relatively weaker connections, to partition the coarsened graph through a novel multi-view graph clustering algorithm.
Finally, in the \textbf{cyclic dependency fixing} (\cref{sec:stage4}),  we address the cyclic dependencies in the clustering results using a heuristic search algorithm and then provide our solution.

\subsection{Code Element Graph Construction}
\label{sec:stage1}
We start with extracting code elements and representing their intricate relationships in a  code element graph.
Here code elements involve definitions/declarations of macros, data types (structs, enums, unions, typedefs), variables, and functions. They cannot be further decomposed.
We extract three types of relationships that hold significance in decomposing a God Header File:
1) Dependency Relationship. It refers to a code element using a name declared by another code element (def-use). 
2) Semantic Relationship. It describes the textual similarity between two code elements.
3) Co-usage Relationship. This relationship characterizes how frequently two code elements are used together in source files and reflects build dependencies.

\sloppy Formally, the code element graph is represented as $G=(V, E^d, E^s, E^c)$, where $V=\{v_i\}_{i=1}^n$ denotes the set of code elements, and $E^d$, $E^s$, $E^c$ denote three types of edges representing dependency, semantic, co-usage relationships respectively. For each edge type $r \in \{d, s, c\}$, $A^r$ is its adjacency matrix and $A^r_{i,j}$ is the weight of edge $(v_i, v_j)$.

We compute the weight of dependency edges following the Call-based Dependence proposed by Bavota et al.~\cite{bavota2011identifying}. Let $Successor_j$ denote the set of code elements using $v_j$, then
\begin{equation}
    d_{i \rightarrow j} = 
        \begin{cases} 
            \frac{1}{|Successor_j|} \ & {\rm if} \ v_i \ {\rm uses} \ v_j\\
            0   \ & {\rm otherwise}
        \end{cases}
\end{equation}
\begin{equation}
    A^d_{i,j} = max \{ d_{i \rightarrow j}, d_{j \rightarrow i}\}
\end{equation}
Note that dependency edges are directional, yet the adjacency matrix is intentionally designed to be symmetrical, ensuring its commutativity.
If $A^d_{i,j}=1$, either $v_j$ is solely used by $v_i$ or vice versa, indicating that they should belong to the same sub-header file.

To calculate the weight of semantic edges, we tokenize and lemmatize identifiers of each code element $v_i$, and then filter out stop words such as ``set'', ``get'', etc, resulting in a word set $Word_i$. These words imply the software concepts associated with the code element. Thus, the semantic similarity between two code elements is calculated as follows:
\begin{equation}
    A^s_{i,j} = \frac{|Word_i \cap Word_j|}{|Word_i \cup Word_j|}
\end{equation}

Regarding co-usage edges, let $File_i$ denote the set of source files using code element $v_i$, the weight between code element $v_i$ and $v_j$ is computed by:
\begin{equation}
    A^c_{i,j} = \frac{|File_i \cap File_j|}{|File_i \cup File_j|}
\end{equation}
If the co-usage weight between two code elements is large, they are more likely to be closely related in functionality. 


\subsection{Dependency-Based Graph Coarsening}
\label{sec:stage2}

As mentioned before, dependency relationships are usually more sparse but inherently convey stronger connections than the other two types of relationships. 
Intuitively, if two code elements exclusively depend on each other, they should be placed within the same sub-header file. 
However, when a code element is dependent upon many other code elements, it is less certain to reach the same conclusion. 
Thus, in this phase, we coarsen the code element graph by merging code elements exhibiting high dependency weights, ensuring the tightly connected code elements remain together throughout subsequent procedures.

Particularly, we iteratively merge two nodes $v_i$ and $v_j$ as a new node $v_{i^{'}}$, if and only if the dependency weight $A^d_{i,j}$ is 1. We set the threshold as 1 to ensure that no mistakes are introduced in this step.
Then we update the weights between the newly formed node $v_{i^{'}}$ and other nodes by $A^d_{i^{'},k} = A^d_{k,i^{'}} = max(A^d_{i,k} \ ,\  A^d_{j,k})$, where $k \neq i$ and $k \neq j$.
The process continues until all nodes with high dependency weights have been merged.

The outcome is a coarsened graph, denoted as $G^{'}$. Each node $v_{i^{'}}$ in it represents either an individual code element or a set of tightly connected code elements.
Additionally, in the adjacency matrices of the coarsened graph, each value ${A^r_{i^{'}, j^{'}}}$ is assigned the maximum weight between code elements from node $v_{i^{'}}$ and node $v_{j^{'}}$.


\subsection{Multi-View Graph Clustering}
\label{sec:stage3}

In this phase, we cluster the coarsened graph to group code elements that share similar functionality while separating those with weaker associations. To achieve this goal, we utilize semantic and co-usage edges as they are dense and encompass latent functional features.

To carry out the clustering task, we apply a novel multi-view graph clustering algorithm DuaLGR~\cite{ling2023dual} to the coarsened graph. DuaLGR takes a shared feature and adjacency matrices of different graph views as input, and refines them by extracting high-level view-common information. 
Furthermore, it adaptively assigns weights and orders to various views to aggregate them into a consensus graph. Then, each node is embedded according to the consensus graph through a GNN-based encoder, facilitating subsequent clustering.

Compared to other clustering algorithms, DuaLGR proves highly suitable for decomposing related software refactoring tasks for several reasons: First, software graphs often encompass non-homophilous edges, as entities from different modules can have connections. This algorithm performs well in both homophilous and low homophilous graphs thanks to its refinement process. Second, it addresses the inconsistency in optimal weights across projects~\cite{bavota2014automating} by dynamically assigning weights and orders in the fusion process. Last, its scalability allows easy incorporation of additional relationships, such as co-evolution, if deemed necessary.

Specifically, we use adjacency matrices $A^s$ and $A^c$ in this stage and concatenate them to construct the shared feature $X$. To refine the adjacency matrices, two labels are introduced. The first one is a soft label $Z_f$, derived from a pretrained autoencoder. This label encapsulates high-level semantic information from both shared feature $X$ and adjacency matrices $A^s$, $A^c$. Then a refinement matrix is calculated as $\Omega = Z_f Z_f^T$.
The second label, a pseudo label, is obtained from the current clustering outcome. It serves to evaluate homophilous rate for each graph view and assign an order $od^r$ based on the value. Notably, graph views with higher homophilous rates will be assigned a higher order.
Each graph view is refined as:
\begin{equation}
    \overline{A}^r = \alpha (\frac{1}{od^r} \sum_{i=1}^{od^r}(A^r)^i) + \Omega 
\end{equation}
where $\alpha$ is a hyperparameter that controls the influence of the homophily of the graph across different views. The refined adjacency matrix contains both high-order structural information and global common information.

To fuse graph views, DuaLGR utilizes the pseudo label to calculate a weight $w^r$ for each view. Finally, the global consensus graph $A$ can be obtained by 
\begin{equation}
    \vspace{-6pt}
    A = w^s \overline{A}^s + w^c \overline{A}^c 
\end{equation}
The consensus graph is then encoded by a Graph Convolution Network~\cite{kipf2016semi} and used for clustering with the K-means algorithm~\cite{lloyd1982least}.


After the clustering process, we acquire several clusters of the coarsened graph. If a node in the coarsened graph represents multiple code elements in the original code element graph, all of them are considered to be in the same cluster to which the node belongs. 

\subsection{Cyclic Dependency Fixing}
\label{sec:stage4}

In the previous stage, we obtain a set of disjoint clusters of code elements $C=\{C_1, \cdots, C_K\}$, each representing a sub-header file. If any code element in cluster $C_i$ relies on a code element in cluster $C_j$ (forming a dependency relationship), an inclusion relationship is established between $C_i$ and $C_j$. If such inclusion relationships forms cyclic dependencies, the decomposition result is unacceptable in our task.
Therefore, our focus during this stage is on the resolution of cyclic dependencies.

\normalem
\begin{algorithm}[!t]
    \caption{$FixCycles(\{C_1, \cdots, C_K\})$}
    \label{alg:fix}
    \KwIn {clustering results $\{C_1, \cdots, C_K\}$ with cycles}
    \KwOut {updated clusters $\{C_1, \cdots, C_K\}$ without cycles}

    \While {there exists cycles in $\{C_1, \cdots, C_K\}$}
    {
        $\{C_{i_1}, \cdots, C_{i_l}\} \leftarrow FindLongestCycle(\{C_1, \cdots, C_K\})$ \; 

        \If {$|\{C_{i_1}, \cdots, C_{i_l}\}| == 2$}
        {
            $C_{i_1}, C_{i_2} \leftarrow FixTwoNodeCycle(C_{i_1}, C_{i_2})$ \;
        }
        \Else 
        {
            $best\_clusters \leftarrow \emptyset$; $best\_gain \leftarrow -\infty$ \; 
            \ForEach {$C_{i_j} \in \{C_{i_1}, \cdots, C_{i_l}\}$}
            {
                $C_{else} \leftarrow \bigcup \{C_{i_k} | k \neq j\}$\;
                $C_{i_j}^{'}, C_{else}^{'} \leftarrow FixTwoNodeCycle(C_{i_j}, C_{else})$ \;
                \ForEach{$C_{i_r} \in \{C_{i_k} | k \neq j\}$}
                {
                    $C_{i_r}^{'} \leftarrow MoveCodeElements(C_{i_r}, C_{i_j}, C_{i_j}^{'})$ \;
                }
                $gain \leftarrow MovingGain(\{C_{i_1}, \cdots, C_{i_l}\}, \{C_{i_1}^{'}, \cdots, C_{i_l}^{'}\})$ \;
                \If {$gain > best\_gain$}
                {
                    $best\_clusters \leftarrow \{C_{i_1}^{'}, \cdots, C_{i_l}^{'}\}$ \;
                    $best\_gain \leftarrow gain$ \;
                }
            }
            $\{C_{i_1}, \cdots, C_{i_l}\} \leftarrow best\_clusters$\;
        }
    }
    \Return $\{C_1, \cdots, C_K\}$ \;

\end{algorithm}

\ULforem

To fix cyclic dependencies, we propose a heuristic algorithm that extends from a two-node cycle fixing method proposed by Herrmann et al.~\cite{herrmann2019multilevel}. 
When dealing with a cycle of length two, which involves $C_i$ and $C_j$, there are two potential directions. We could either eliminate the inclusion from $C_i$ to $C_j$ or that from $C_j$ to $C_i$. In the former scenario, one option is to move all ancestors of the nodes in $C_i$ to $C_i$, ensuring that there are no edges from $C_i$ to $C_j$. Alternatively, we can also move all descendants of the nodes in $C_j$ to $C_j$, which also guarantees acyclicity. These two operations can similarly be applied to eliminate the inclusion from $C_j$ to $C_i$. 
With four possible choices to rectify a two-node cycle, we calculate the moving gain for each choice and select the optimal one. The moving gain is designed based on the moved nodes and dependency edges. Let $V_{m}$ denote the set of nodes moving from $C_i$ to $C_j$, the moving gain is calculated as follows:

\begin{footnotesize}
\begin{equation}
\begin{split}
\label{equ:gain}
    gain = \sum_{\substack{v_r \in V_{m}, \\ v_s \in C_{j}} } (A^d_{r,s} + A^d_{s,r}) 
                    - \sum_{\substack{v_r \in V_{m}, \\ v_s \in C_{i}-V_{m}} }(A^d_{r,s} + A^d_{s,r})
                    - |V_{m}|
\end{split}
\end{equation}
\end{footnotesize}
It calculates the summation of all dependency weights between $V_{m}$ and $C_i$,  subtracting the summation of all dependency weights between $V_{m}$ and $C_j$. Besides, we hope to move as few nodes as possible, so we also subtract the count of moved nodes in the formula.

To handle cycles longer than two, we adopt a strategy of reducing them to two-node cycles. The details of this process are illustrated in Algorithm~\ref{alg:fix}. 
For a cycle with length $l > 2$, we pick one cluster $C_{i_j}$ from the cycle and merge other clusters into a single cluster (line 8). Then we apply the two-node cycle fixing method described above (line 9). Based on the outcome, we update each cluster in the cycle (line 11): if some code elements are moved to $C_{i_j}$, they are removed from their original clusters; conversely, if certain code elements are moved from $C_{i_j}$, they are assigned to the cluster with the highest dependency weight. We traverse all possible choices and select the solution with the highest moving gain.

We iteratively choose the longest cycle (line 2) and reduce its length by one on each iteration. While this operation may introduce new shorter cycles, it will never introduce a cycle of equal or longer length. And if the longest cycle has a length of 2, no new cycles will be introduced. Consequently, our approach will eventually terminate with all cycles eliminated.

The worst-case time complexity of our heuristic search algorithm is $O(2^n)$. However, in our specific scenario, the number of sub-header files derived from a God Header File is relatively small in practice. Consequently, the iteration times remain manageable, ensuring that the cost of our approach is acceptable.


\section{Evaluation Setup}
\label{sec:evaluation}

In this section, we present our experimental methodology and evaluation setup. 
Our exploration is guided by the following research questions:

\begin{itemize}[left=2em]
    \item[\textbf{RQ1.}] \textbf{How well does our approach perform in terms of accuracy?}

    In this RQ, we investigate whether existing God Class refactoring methods are applicable for the task of decomposing God Header Files and to what extent our approach enhances the accuracy.
     
    \item[\textbf{RQ2.}] \textbf{How well does our approach perform in terms of refining architecture?}

    This question helps us understand the architectural benefits of our approach in decomposing real-world God Header Files.
    
    \item[\textbf{RQ3.}] \textbf{How well does our approach perform in terms of reducing recompilation? }

    This question helps us understand the extent to which our decomposition results could reduce recompilation based on the historical commits.
    
\end{itemize}

\subsection{Datasets}

To evaluate the effectiveness of our approach, we have constructed two datasets. One is a synthetic dataset containing artificially created God Header Files, while the other is derived from real-world open-source projects. The datasets are available in~\cite{web:dataset}.

\textbf{Synthetic God Header Files.} To evaluate the accuracy of our approach, we followed Bavota et al.~\cite{bavota2014automating} and artificially created 9 God Header Files by merging several smaller header files. Specifically, we selected 8 header files respectively from three open-source projects: PolarDB-for-PostgreSQL~\cite{PolarDB-for-PostgreSQL} (abbreviated as PolarDB), fontforge~\cite{fontforge}, and FreeRDP~\cite{FreeRDP}, then merged 4/6/8 of them to assemble each God Header File. 
Notably, we deliberately chose header files whose cohesion is higher than the average cohesion of the project. This criterion is employed to filter out poorly designed header files and guarantee the quality of the ground truth.


\textbf{Real-world God Header Files.}
To assess the practical performance of our approach, we also employed it on the typical God Header Files identified in section~\ref{subsec:preliminary} (Table~\ref{projects_and_headerfiles}). These files, chosen from manually confirmed God Header Files, possess the highest number of commits in their respective projects. 
They are characterized by an extensive amount of code elements, influencing hundreds of thousands of lines of code. The abundant commit histories of God Header Files is advantageous for us to estimate the reduction in recompilation after decomposition.



\subsection{Comparison Methods}

Since we are the first to address the issue of God Header File, we selected three prior state-of-the-art God Class refactoring methods as our comparison methods due to the similarity between these two problems:

\begin{itemize} 
    \item \textbf{Bavota et al.}~\cite{bavota2014automating} analyze relationships between the methods in a class to identify chains of strongly related methods, which are used to define new classes with higher cohesion than the original class.
    \item \textbf{Wang et al.}~\cite{wang2017automatic} describe a system-level refactoring algorithm that can identify multiple refactoring opportunities automatically. For class extraction, a multi-relation network is constructed and a weighted clustering algorithm is applied for regrouping methods.
    \item \textbf{Akash et al.}~\cite{akash2022exploring} utilize graph auto-encoder for learning a vector representation for each method (as node) in the class after constructing an initial graph. The learned vectors are used to cluster methods into different groups to be recommended as refactored classes.
\end{itemize}

We chose Bavota's~\cite{bavota2014automating} and Wang's~\cite{wang2017automatic} approaches for comparison as they represent the state of the art. We did not include the well-known JDeodorant~\cite{fokaefs2011jdeodorant, fokaefs2012identification} because both Bavota's and Wang's studies claimed superior performance compared to it. Additionally, Akash's~\cite{akash2022exploring} approach was selected due to its use of Graph Neural Networks (GNNs), aligning with our own approach which also employs a GNN-based clustering algorithm.

In order to apply these methods to the task of God Header File decomposition, we reproduced their code similarity calculation module. In this process, we adhere to the original design of each comparison method, substituting all references to ``attribute'' and ``method'' with ``code element'', and replace attribute access and method invocation with define-use relationships between code elements.
Then we adopt the settings of clustering algorithms exactly as described in the comparison works.


\subsection{Metrics}


To assess the accuracy of our approach and answer RQ1, we measured each generated decomposition result based on its closeness to the expected result using a set of common metrics, which are described below. 

\begin{itemize} 
    \item \textbf{MoJoFM}~\cite{wen2004effectiveness} quantifies the number of Move and Join operations required to transform architecture A into B. It is calculated as:
    \begin{equation}
    \small
        MoJoFM(A,B)= (1-\frac{mno(A,B)}{\max(mno(\forall A,B))}) \times 100\% 
    \end{equation}
    where $mno(A,B)$ is the minimum number of \textit{Move} or \textit{Join} operations needed to transform the partition A to B. MoJoFM scores range from 0\% to 100\%, where a higher value indicates a higher similarity between two partitions.

    \item \textbf{Normalized Mutual Information (NMI)}~\cite{mcdaid2011normalized} measures the mutual information between the predicted clusters and ground truth clusters. It normalizes the mutual information to fall within a range of 0 to 1, with 1 indicating perfect clustering agreement.
    
    \item \textbf{Adjust Rand Index (ARI)}~\cite{yeung2001details} is based on the pairwise similarity between predicted labels and ground truth labels. It adjusts for random chance, providing a normalized score between -1 and 1, where higher values indicate better clustering.

    \item \textbf{Accuracy (ACC)}~\cite{lutov2019accuracy} measures the proportion of correctly predicted cluster labels by building a mapping between predicted cluster labels and ground truth cluster labels through the Kuhn-Munkres~\cite{lovasz2009matching} algorithm.

    \item \textbf{F1-score (F1)} is also a metric for evaluating the clustering performance, which combines precision and recall.
\end{itemize}



To evaluate the performance on the real-world dataset and answer RQ2 and RQ3, we used the following metrics to assess the architecture and reduction in recompilation of the decomposition results.

\begin{table*}[!htbp]
\centering
\caption{Performance of different decomposing methods on synthetic God Header Files.}
\resizebox{\linewidth}{!}{
    \begin{tabular}{l|ccccc|ccccc|ccccc}
    \toprule[1.5pt]
    
        \multirow{2}{*}{Methods/Datasets}                
        & \multicolumn{5}{c|}{PolarDB (\#subfiles=4 \& \#code elements=138)}      
        & \multicolumn{5}{c|}{PolarDB (\#subfiles=6 \& \#code elements=276)}        
        & \multicolumn{5}{c}{PolarDB (\#subfiles=8 \& \#code elements=362)}  \\
        
        & \multicolumn{1}{c}{MoJoFM} & \multicolumn{1}{c}{NMI(\%)} & \multicolumn{1}{c}{ARI(\%)} & \multicolumn{1}{c}{ACC(\%)} & \multicolumn{1}{c|}{F1(\%)}
        & \multicolumn{1}{c}{MoJoFM} & \multicolumn{1}{c}{NMI(\%)} & \multicolumn{1}{c}{ARI(\%)} & \multicolumn{1}{c}{ACC(\%)} & \multicolumn{1}{c|}{F1(\%)}
        & \multicolumn{1}{c}{MoJoFM} & \multicolumn{1}{c}{NMI(\%)} & \multicolumn{1}{c}{ARI(\%)} & \multicolumn{1}{c}{ACC(\%)} & \multicolumn{1}{c}{F1(\%)} \\
        
    \midrule

        Bavota et.al.   &69.9   &63.0   &50.5   &64.2   &61.0       
                        &56.1   &44.2   &20.9   &56.7   &36.2       
                        &42.9   &51.7   &32.1   &42.9   &15.8   \\
        Wang et.al.     &96.2   &90.6   &90.9   &96.4   &96.2       
                        &87.4   &90.2   &89.7   &87.3   &76.0       
                        &87.9   &87.6   &87.1   &87.8   &79.4   \\
        Akash et.al.    &75.9   &59.2   &51.4   &76.6   &75.2       
                        &75.8   &63.2   &73.1   &76.0   &66.9       
                        &68.8   &61.3   &66.8   &66.8   &58.0   \\
        \textbf{Ours}            &\textbf{100.0}  &\textbf{100.0}  &\textbf{100.0}  &\textbf{100.0}  &\textbf{100.0}       
                        &\textbf{89.2}   &\textbf{89.3}   &\textbf{92.1}   &\textbf{89.5}   &\textbf{82.6}       
                        &\textbf{90.1}   &\textbf{91.4}   &\textbf{90.8}   &\textbf{90.0}   &81.7  \\
    
        w/o coarsening      &96.2   &92.0   &92.7   &97.1   &96.9       
                            &88.1   &88.1   &90.0   &87.3   &77.9       
                            &89.5   &88.2   &86.6   &89.8   &\textbf{85.4}   \\
        w/o multi-view  &93.9   &87.0   &87.0   &94.9   &94.8       
                            &87.7   &87.3   &78.2   &81.8   &75.4       
                            &88.1   &88.5   &78.7   &83.9   &79.8   \\
    \midrule[1pt]

        Methods/Datasets                
        & \multicolumn{5}{c|}{fontforge (\#subfiles=4 \& \#code elements=145)}      
        & \multicolumn{5}{c|}{fontforge (\#subfiles=6 \& \#code elements=292)}        
        & \multicolumn{5}{c}{fontforge (\#subfiles=8 \& \#code elements=342)}  \\

    \midrule

        Bavota et.al.   &41.4   &13.1   &0.4    &38.2   &31.6       
                        &51.6   &25.9   &26.4   &45.7   &30.2       
                        &45.9   &23.5   &15.1   &40.8   &26.0   \\
        Wang et.al.     &97.9   &95.2   &96.2   &98.6   &98.5       
                        &71.9   &65.2   &47.3   &71.5   &66.0       
                        &65.0   &61.9   &41.6   &64.5   &56.2   \\
        Akash et.al.    &53.6   &24.0   &14.8   &52.1   &49.9       
                        &53.7   &28.7   &20.6   &49.5   &46.2       
                        &49.5   &29.1   &18.1   &45.2   &41.5   \\
        \textbf{Ours}            &\textbf{100.0}  &\textbf{100.0}  &\textbf{100.0}  &\textbf{100.0}  &\textbf{100.0}       
                        &\textbf{89.1}   &\textbf{82.8}   &\textbf{75.7}   &\textbf{89.6}   &\textbf{89.1}      
                        &\textbf{88.0}   &\textbf{82.8}   &\textbf{72.9}   &\textbf{88.6}   &\textbf{88.8}   \\
        w/o coarsening      &99.3   &97.8   &98.8   &99.3   &98.9       
                            &78.9   &68.4   &50.4   &73.9   &72.7      
                            &80.8   &72.8   &54.3   &78.0   &79.8   \\
        w/o multi-view  &100.0   &100.0   &100.0   &100.0   &100.0       
                            &75.0   &67.7   &46.4   &68.0   &62.5       
                            &87.6   &81.3   &71.1   &88.3   &88.6   \\                       

    \midrule[1pt]

        Methods/Datasets                
        & \multicolumn{5}{c|}{FreeRDP (\#subfiles=4 \& \#code elements=217)}      
        & \multicolumn{5}{c|}{FreeRDP (\#subfiles=6 \& \#code elements=397)}        
        & \multicolumn{5}{c}{FreeRDP (\#subfiles=8 \& \#code elements=534)}  \\

    \midrule

        Bavota et.al.   &87.2   &71.5   &57.2   &65.6   &58.9       
                        &91.2   &85.0   &85.0   &78.9   &68.2       
                        &91.8   &85.2   &80.1   &73.2   &61.3   \\
        Wang et.al.     &90.5   &79.4   &61.3   &68.4   &63.6       
                        &80.2   &76.6   &66.5   &71.8   &42.6       
                        &81.0   &79.5   &67.7   &72.1   &41.6   \\
        Akash et.al.    &89.1   &74.3   &60.2   &67.0   &61.2       
                        &88.4   &78.9   &80.7   &76.1   &66.6       
                        &80.3   &68.3   &67.6   &65.1   &55.7   \\
        \textbf{Ours}            &\textbf{99.1}   &\textbf{96.3}   &\textbf{97.8}   &\textbf{99.1}   &\textbf{98.8}      
                        &\textbf{99.2}   &\textbf{98.3}   &\textbf{98.9}   &\textbf{99.5}   &\textbf{99.3}       
                        &\textbf{99.0}   &\textbf{98.5}   &\textbf{99.1}   &\textbf{99.4}   &\textbf{98.9}   \\
        w/o coarsening      &98.6   &95.4   &98.0   &98.6   &97.3       
                            &98.7   &96.8   &98.5   &99.0   &98.0       
                            &98.9   &97.5   &98.6   &99.1   &98.3   \\
        w/o multi-view  &67.7   &50.9   &27.7   &59.6   &56.1       
                            &81.1   &69.1   &53.4   &81.7   &84.3       
                            &68.1   &55.5   &27.4   &59.6   &56.1   \\    
    
    \bottomrule[1.5pt]
    \end{tabular}
}
\label{tab:accuracy}
\vspace{-5pt}
\end{table*}

\begin{itemize} 
    \item \textbf{Modularity}~\cite{newman2004finding} is a commonly used metric to evaluate the quality of a graph's partition. It measures the extent to which a graph can be partitioned into clusters with more connections within the clusters than would be expected if the connections were randomly distributed. It is calculated by:
    \begin{equation}
        Q = \frac{1}{2m}\sum_{i,j}[A_{ij}-\frac{k_ik_j}{2m}]\delta (c_i,c_j)
    \end{equation}
    where $A_{ij}$ represents the weight of the edge between node $v_i$ and node $v_j$, $k_i=\sum_j A_{ij}$ is the sum of weights of all edges connected to node $v_i$, $m$ is the sum of weights of all edges in the graph, and $\delta(c_i,c_j)$ is 1 if $v_i$ and $v_j$ are in the same cluster, 0 otherwise. In our code element graph, the weight of an edge between two nodes is calculated as: $A_{ij}=A^s_{i,j}+A^c_{i,j}+A^d_{i,j}$. A higher modularity score indicates a better architectural design.

    \item \textbf{Recompilation Cost}. We assessed the recompilation cost of each decomposition result by examining the commit history of the original God Header File. 
    By analyzing the build dependencies, we are able to identify the file set need to be recompiled under a specific commit. Then we calculate the number of files and the lines of code to be recompiled. To estimate the recompilation time, we followed McIntosh et al.~\cite{mcintosh2016identifying}, and recorded the elapsed time spent compiling each translation unit.
\end{itemize}

\section{Evaluation Results}
\label{sec:results}

We now describe our evaluation results for each research question.

\subsection{RQ1: Accuracy}

In RQ1, we applied the three comparison methods along with our approach to the synthetic dataset. Since there is an expected number of subfiles for each synthetic God Header File to decompose, we adjusted the hyperparameters of each method to ensure that the decomposition results produce precisely that number of subfiles.
The performance of each decomposing method and the results of our approach under each setting is presented in Table~\ref{tab:accuracy}.

Notably, our approach outperforms the three God Class refactoring methods when decomposing synthetic God Header Files.
On one hand, our approach achieves the \textbf{highest accuracy} across all synthetic files and all metrics. Compared with Wang's approach, which exhibits the best performance among the comparison methods, our approach improves the MoJoFM by 11.5\% on average. 
On the other hand, our approach demonstrates \textbf{enhanced consistency} across different software projects. In contrast, the approaches of Bavota and Akarsh exhibit considerable variation in their performance on different software projects. They achieve high accuracy, with over 80 MoJoFM, in the FreeRDP project but lower accuracy, approximately 50 MoJoFM, in the fontforge project. And Wang's approach outperforms the other two comparison methods in the case of the PolarDB and fontforge projects but falls behind in the FreeRDP project. The reason for the general superiority of Wang's approach lies in its consideration of the functional coupling weight (FCW), which is similar to the co-usage edges in our approach. However, its relatively worse performance in the FreeRDP project can be attributed to its reliance on a set of fixed weights to aggregate all code relationships, which do not align well with all projects.

Our approach addresses the above issue by incorporating a novel multi-view graph clustering algorithm. It dynamically assigns weights tailored to different projects, resulting in consistent performance across various software projects. The effectiveness of this multi-view graph clustering component is evident from the results of our ablation study (w/o multi-view). When this component is replaced with k-means, which takes an average of semantic adjacency and co-usage adjacency as input, the performance of the approach declines on almost all synthetic files. 
The performance decrease is relatively slight on PolarDB and fontforge, but quite significant on FreeRDP, approximately 30\%. This discrepancy comes from the varying importance of different code relationships for different projects. In such cases, using fixed weights to combine multiple code relationships is not appropriate, underscoring the superiority of the mechanism of dynamically assigning weights in multi-view graph clustering.

Meanwhile, the absence of graph coarsening component also leads to a decrease in the performance (w/o coarsening), highlighting its contribution to the effectiveness of our approach. This decrease is rather subtle for most of the synthetic files, typically below 5\%. But the decline is more significant for two synthetic files consisting of 6 and 8 files from fontforge. The reason is that the percentage of dependency edges is higher in these files compared to the others, making the graph coarsening component particularly essential for such files.

\begin{RQResult}
{Summary for RQ1}{In comparison to existing methods, our approach achieves 11.5\% higher accuracy on all synthetic files and demonstrates enhanced consistency across various software projects. Both the graph coarsening component and the multi-view graph clustering component make a notable contribution to the performance of our approach.
}
\end{RQResult}

\subsection{RQ2: Architectural benefits}

We also employed the three God Class refactoring methods and our approach to the six real-world God Header Files to investigate the performance. In this situation, the number of subfiles required for decomposition is uncertain, so we empirically configured the number within a range from 4 to 10 for all six files , as developers won't prefer a larger number in practice.
In this section, we will discuss the architectural benefits of our approach from the perspective of acyclic dependencies and higher modularity among the sub-header files of the decomposition results.

\begin{figure*}[!htbp]
    \centering
    \includegraphics[width=\linewidth]{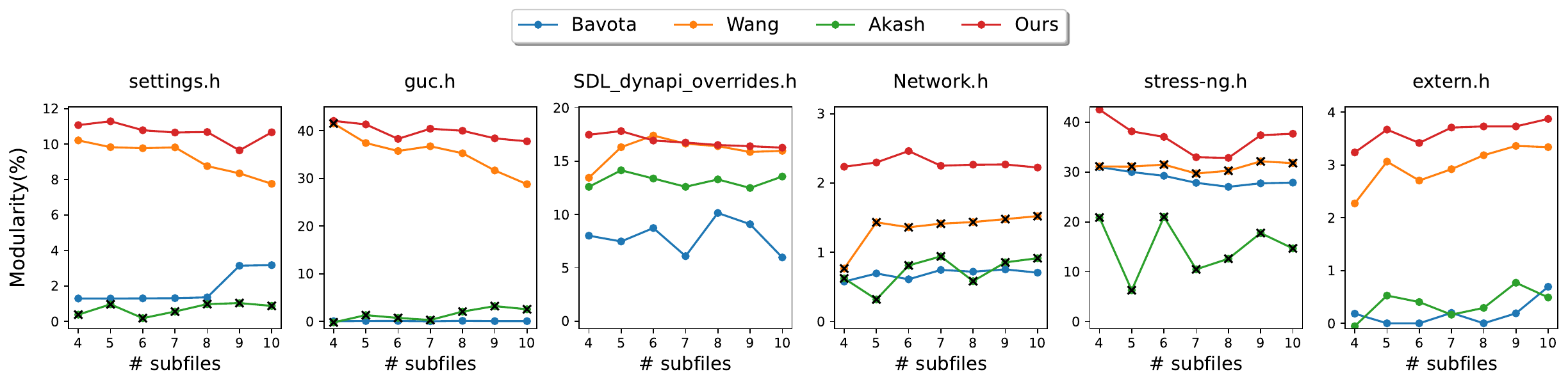}
    \vspace{-5pt}
    \caption{Modularity of different methods on real-world God Header Files. Black cross markers represent the decomposition results with cyclic dependencies.}
    \label{fig:real-world_results}
    \vspace{-5pt}
\end{figure*}

In Figure~\ref{fig:real-world_results}, we calculated modularity for each decomposition result and plotted the values as functions of the number of subfiles for different God Header Files. The decomposition results that suffered cyclic dependencies, which may lead to compilation errors, are marked by black cross markers.
In our experiment, each refactoring method is applied to 6 files across 7 different settings, totaling 42 cases. Among these, 15 results of Wang's approach and 28 of Akash's exhibited cyclic dependencies, rendering them infeasible. Bavota's approach, in contrast, produced results without cycles due to its connected subgraph based clustering algorithm. However, these results often performed poorly on modularity.
As for our approach, the cyclic dependency fixing module ensures acyclic dependencies among the decomposed sub-header files, and the multi-view graph clustering module guarantees high-level modularity, resulting in a simpler and clearer architecture.

To be specific, our approach consistently yields the highest modularity across almost all configurations, with an average improvement of 9.1\% compared to the best comparison method. This indicates that our approach generates sub-header files that adhere closely to the principles of high cohesion and low coupling, thereby facilitating architectural redesign.
For the file \texttt{SDL\_dynapi\_overrides.h}, Wang's approach achieves comparable modularity with ours. This is because only one type of relationship dominates in this file: there are no dependencies and the number of co-usage edges is an order of magnitude less than semantic edges. In this scenario, the multi-view graph clustering module of our approach could not leverage its advantage.
Meanwhile, the values of modularity exhibit significant variations among different files. For instance, \texttt{guc.h} achieves a remarkable 40\%, while \texttt{Network.h} struggles to surpass 3\%. This discrepancy could be attributed to the inherent complexity and interwoven nature of the code elements in \texttt{Network.h}. Such God Header Files are possibly beyond the capabilities of automatic refactoring methods, emphasizing the necessity of involving experts in the architectural redesign.


\begin{figure}[!tbp]
    \centering
    \begin{subfigure}{0.23\textwidth}
        \includegraphics[width=\textwidth]{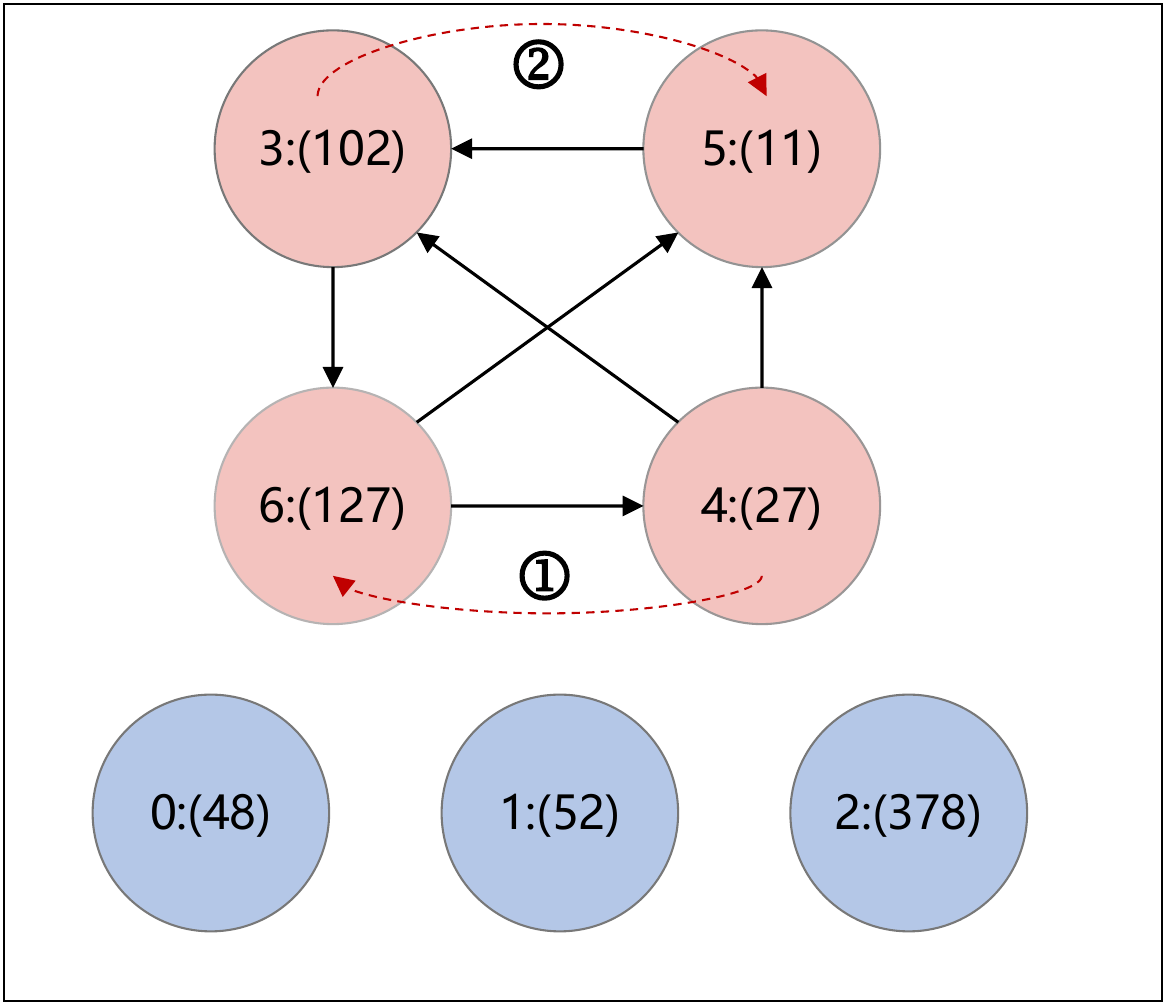}
        \caption{Fixing process.}
        \label{fig:unfixing}
    \end{subfigure}
    \begin{subfigure}{0.23\textwidth}
        \includegraphics[width=\textwidth]{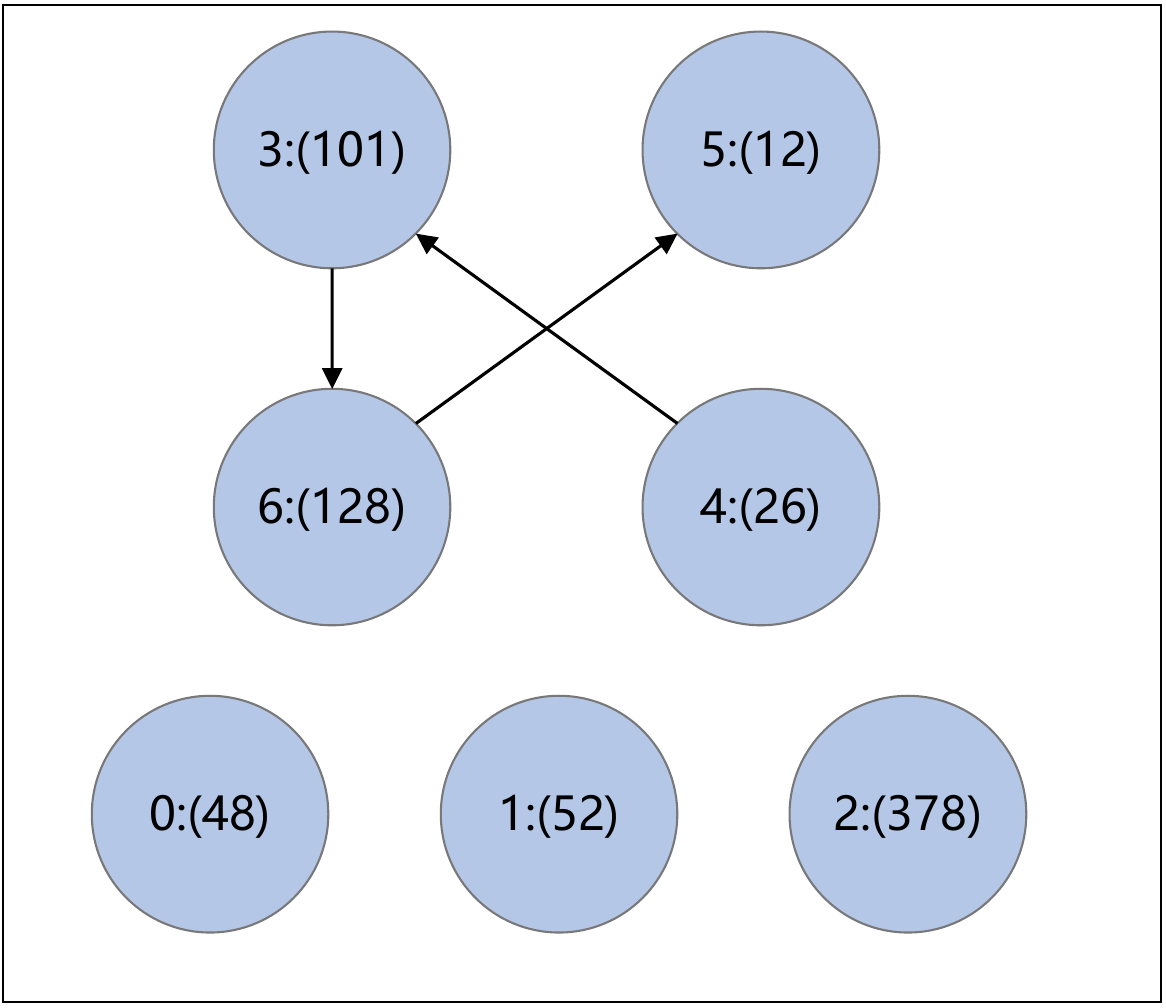}
        \caption{Fixing result.}
        \label{fig:fixing2}
    \end{subfigure}
  \caption{The process of cyclic dependency fixing of settings.h(\#subfiles=7). Each circle represents a subfile with a label $i:(|C_i|)$. Each arrow represents an include relationship. And the subfiles involved in cycles are marked with color red. }
  \label{fig:case}
  \vspace{-5pt}
\end{figure}



To illustrate the cyclic dependency fixing process, we provide an example in Figure~\ref{fig:case}.
Figure~\ref{fig:unfixing} presents the clustering result for \texttt{settings.h} (\#subfiles=7) after the multi-view graph clustering phase. In this instance, there are three cycles: 3-6-4-5, 3-6-4, and 3-6-5. Following our cyclic dependency fixing algorithm, we prioritize addressing the longest cycle, which is 3-6-4-5. The heuristic search process yields the optimal solution: moving the code element ``rdpSettings'' from subfile 4 to subfile 6. This movement involves only one code element and attains the highest gain of 61, eliminating dependencies from 4 to 5 and from 6 to 4. As a result, subfile 4 is no longer involved in any cycles. 
The intermediate result after this movement contains only one cycle: 3-6-5. In the next step, the best move is to relocate the code element ``ALIGN64'' from 3 to 5 based on the gain of each choice. Then all cycles are eradicated, and the final clustering result is presented in Figure~\ref{fig:fixing2}.

\begin{RQResult}
{Summary for RQ2}{Our decomposed sub-header files exhibit superior architecture. They are characterized by an average improvement of 9.1\% on modularity compared to the state-of-the-art comparison method, along with simplified acyclic dependencies.
}
\end{RQResult}


\subsection{RQ3: Reduction in recompilation}

In this section, we aim to investigate the extent to which the decomposition results could reduce recompilation.
To estimate recompilation cost, we collected the commit history for each real-world God Header File and extracted the modified code elements of each commit. 
For a God Header File, all the files including it require recompilation once it is modified. Through decomposition, a commit that modifies the original header file may affect only a portion of the sub-header files. In this case, code files that only include the unmodified sub-header files do not need to be recompiled.
By analyzing the build dependencies, we are able to identify the file set need to be recompiled under a specific commit. Then we calculate the number of files and the lines of code to be recompiled.
In practice, it takes much effort to calculate the actual recompilation time given a code commit. With many commits that have modified the header file, it is not a trivial task to redo all these modifications (especially in an intermediate version of the system). Therefore, we followed McIntosh et al.~\cite{mcintosh2016identifying}, and recorded the elapsed time spent compiling each translation unit, based on which, we could estimate how much time it takes to recompile under a given commit.

Table~\ref{tab:best_version} shows the average recompilation cost per commit of the original God Header Files as well as the decomposition results generated by each approach. For each approach, we only present the \#subfiles setting on which the approach achieves lowest recompilation cost.
Generally, the decomposition results produced by our approach have the potential to reduce recompilation by a significant margin, ranging from 15\% to 60\%.
And the recompilation cost of our decomposition results is always the lowest except for \texttt{guc.h}.
This happens due to the data skew in the commits of \texttt{guc.h}. This file only has 61 commits and 58 of them did not modify the largest subfile of Wang's result.
However, although Wang's approach reduce more recompilation cost on \texttt{guc.h}, it failed to generate feasible results for \texttt{Network.h} and \texttt{stress-ng.h} due to the issue of cyclic dependencies.
On the other hand, Bavota's approach successfully decomposed all six files but consistently demonstrated lower performance in terms of recompilation reduction. Only our approach achieves both consistent successful decomposition and substantial reduction in recompilation.

\begin{table*}[!htbp]
\centering
\caption{Average recompilation cost per commit and total saved time of decomposition results.}
\resizebox{\linewidth}{!}{
\begin{tabular}{l|ccccc|ccccc}
    \toprule[1pt]
    
        \multirow{2}{*}{Methods/Datasets}                
        & \multicolumn{5}{c|}{settings.h}      
        & \multicolumn{5}{c}{guc.h}   \\

        & \multicolumn{1}{c}{\#subfiles}  & \multicolumn{1}{c}{\#recompiled files} & \multicolumn{1}{c}{recompiled LOC} & \multicolumn{1}{c}{\makecell[c]{recompilation\\time(seconds)}} & \multicolumn{1}{c|}{\makecell[c]{total saved\\time(minutes)}}
        & \multicolumn{1}{c}{\#subfiles}  & \multicolumn{1}{c}{\#recompiled files} & \multicolumn{1}{c}{recompiled LOC} & \multicolumn{1}{c}{\makecell[c]{recompilation\\time(seconds)}} & \multicolumn{1}{c}{\makecell[c]{total saved\\time(minutes)}}  \\
        
    \midrule
        before decomposing    & - & 368   & 279,617   & 252.4   & -
                                & - & 388   & 741,265   & 260.3 & -    \\
        Bavota et.al.       & 10     & 250.3($\downarrow$31.9\%)    & 197,735($\downarrow$29.3\%)   & 177.5($\downarrow$29.7\%) & 385.5
                            & 10     & 337.7($\downarrow$13.0\%)    & 648,162($\downarrow$12.6\%)   & 227.2($\downarrow$12.7\%) & 33.7 \\
        Wang et.al.         & 8     & 225.5($\downarrow$38.7\%)    & 180,629($\downarrow$35.4\%)   & 160.6($\downarrow$36.3\%)  & 472.5
                            & 10     & \textbf{214.3}($\downarrow$\textbf{44.8}\%)    & \textbf{432,930}($\downarrow$\textbf{41.6}\%)   & \textbf{147.9}($\downarrow$\textbf{43.2}\%)   & \textbf{114.3} \\
        Akash et.al.        & -     & -    & -   & -    & -
                            & -     & -    & -   & -    & -    \\
        Ours                & 9     & \textbf{163.0}($\downarrow$\textbf{55.7}\%)    & \textbf{145,612}($\downarrow$\textbf{47.9}\%)   & \textbf{130.7}($\downarrow$\textbf{48.2}\%)    & \textbf{626.7}
                            & 7     & 281.5($\downarrow$27.4\%)    & 562,000($\downarrow$24.2\%)   & 190.6($\downarrow$26.8\%)  & 70.8\\
    
    \midrule[1pt]

        & \multicolumn{5}{c|}{SDL\_dynapi\_overrides.h}      
        & \multicolumn{5}{c}{Network.h}   \\ 
    \midrule
        before decomposing    & - & 127   & 100,654   & 215.4   & -
                                & - & 77   & 268,084   & 113.4  & -    \\
        Bavota et.al.       & 9     & 75.9($\downarrow$40.2\%)    & 63,911($\downarrow$36.5\%)   & 133.5($\downarrow$38.0\%)    & 152.9
                            & 10     & 64.8($\downarrow$15.9\%)    & 242,437($\downarrow$9.6\%)   & 101.6($\downarrow$10.4\%)   & 9.8 \\
        Wang et.al.         & 10     & 50.4($\downarrow$60.3\%)    & 50,811($\downarrow$49.5\%)   & 96.0($\downarrow$55.4\%)    & 223.0
                            & -     & -   & -   & - &-  \\
        Akash et.al.        & 9     & 69.5($\downarrow$45.3\%)    & 63,171($\downarrow$37.2\%)   & 128.4($\downarrow$40.4\%)    & 162.4
                            & -     & -    & -   & -    &-    \\
        Ours                & 10     & \textbf{42.9}($\downarrow$\textbf{66.2}\%)    & \textbf{45,594}($\downarrow$\textbf{54.7}\%)   & \textbf{86.2}($\downarrow$\textbf{60.0}\%)  & \textbf{241.1}
                            & 8     & \textbf{56.1}($\downarrow$\textbf{27.1}\%)    & \textbf{210,925}($\downarrow$\textbf{21.3}\%)   & \textbf{88.9}($\downarrow$\textbf{21.6}\%)  & \textbf{20.4}    \\

    \midrule[1pt]

        & \multicolumn{5}{c|}{stress-ng.h}      
        & \multicolumn{5}{c}{extern.h}   \\ 
    \midrule
        before decomposing    & - & 314   & 156,573   & 284.0   &- 
                                & - & 324   & 171,335   & 1047.0    &-    \\
        Bavota et.al.       & 10     & 268.6($\downarrow$14.5\%)    & 139,526($\downarrow$10.9\%)   & 234.0($\downarrow$17.6\%) & 291.2
                            & 10     & 305.4($\downarrow$5.7\%)    & 162,749($\downarrow$5.0\%)   & 987.5($\downarrow$5.7\%)    & 1350.5 \\
        Wang et.al.         & -     & -    & -   & -    &- 
                            & 7     & 283.0($\downarrow$12.6\%)    & 152,740($\downarrow$10.8\%)   & 916.1($\downarrow$12.5\%)  & 2974.3 \\
        Akash et.al.        & -     & -    & -   & -    & -
                            & 9     & 291.4($\downarrow$10.1\%)    & 159,601($\downarrow$6.8\%)   & 943.6($\downarrow$9.9\%)    & 2348.7    \\
        Ours                & 6     & \textbf{256.1}($\downarrow$\textbf{18.4}\%)    & \textbf{135,175}($\downarrow$\textbf{13.7}\%)   & \textbf{222.5}($\downarrow$\textbf{21.6}\%)    & \textbf{358.6}
                            & 9     & \textbf{273.2}($\downarrow$\textbf{15.7}\%)    & \textbf{144,880}($\downarrow$\textbf{15.4}\%)   & \textbf{886.2}($\downarrow$\textbf{15.4}\%)    & \textbf{3653.3}    \\

    \bottomrule[1pt]
\end{tabular}
}
\label{tab:best_version}
\vspace{-5pt}
\end{table*}

Similar to modularity, recompilation reduction also varies across different files. For example, decomposing \texttt{extern.h} only reduces recompilation on an average of 15\%, while this number rises to 60\% for \texttt{SDL\_dynapi\_overrides.h}. Even though the percentage is not very high for \texttt{extern.h}, since it had been modified over 1000 times, our approach could save 60 hours of recompilation during its evolution.
In practice, decomposing God Header Files can yield greater recompilation benefits for  projects with larger build cost and more frequent modifications.


\begin{RQResult}
{Summary for RQ3}{
Decomposing God Header Files into subfiles using our approach can significantly reduce recompilation cost, ranging from 15\% to 60\%. Our method generally achieves the lowest recompilation cost compared to other approaches and can save up to 60 hours for a single God Header File over its historical evolution.
}
\end{RQResult}


\section{Discussion}
\label{sec:discussion}
In this section, we discuss the application, limitations and threats to validity of our approach.

\subsection{Application and Limitation}

Our approach aims to provide an improved decomposition solution for God Header Files. In the experiments, the appropriate number of sub-header files can be suggested by balancing modularity and recompilation reduction.  However, users may feel the recommended number of subfiles a bit excessive in application, as they may prefer a more concise file structure. Therefore, we allow users to assign the number of subfiles when practically applying our approach in their projects. We will also investigate how to suggest the best number automatically in the future.

In our approach, a cycle dependency fixing algorithm is designed to guarantee that dependencies among the decomposed files are acyclic. However, this stage may disrupt the structure of clustering results in case the dependency relationships are complex. In future work, we will investigate how to integrate the acyclic constraint into the clustering process. Although there exists researches on acyclic partition~\cite{moreira2017graph, moreira2017evolutionary, herrmann2019multilevel}, our specific task presents unique challenges, as our graph contains both directed and undirected edges.

The current version of our approach does not try to  assign a good name for the decomposed sub-header file. Ideally, each sub-header file should be provided with a file name or a brief descriptive comment that reflects the concepts and functionalities of the file for developers to comprehend its contents quickly. We will further explore the process of generating appropriate file names and concise descriptive comments for these decomposed sub-header files.

\subsection{Threats to Validity}

Our experiment results may suffer from several threats to validity. We discuss internal validity and external validity respectively.

For \textbf{internal validity}, we have synthesized several God Header Files based on small header files, considering  the original small header files as the ground truth to compute accuracy. However, the ground truth may contain errors if the original header file is poorly designed. We try to mitigate this issue by selecting header files with high cohesion.
In addition, both the comparison methods and our approach employ clustering algorithms that exhibit a certain level of randomness due to randomization in their initialization and aggregation processes. As a result, the clustering results can vary across different runs. To mitigate these potential biases, we conducted three separate runs for each decomposing method and reported the best results.
Another factor potentially affecting our results is the metrics we use to evaluate architecture and recompilation. The results and the conclusions of RQ2 and RQ3 are scoped by the efficacy of the metrics. Moreover, the recompilation cost is calculated based on historical commits, thus decomposition results with lower recompilation cost might not perform better on future commits.

For \textbf{external validity}, we have evaluated our approach on a diverse set of both synthetic and real-world God Header Files. We also employed several God Class refactoring methods for comparison. It is possible that the outcomes vary when applying the methods to a different set of header files. Therefore, exercising caution is essential when generalizing our findings to other files or algorithms.



\section{Related Work}
\label{sec:related_work}

Given the relevance of the tasks and the techniques, we present the related work about header file optimization, God Class refactoring, and graph clustering.

\subsection{Header File Optimization}

There are plenty of researches aiming at reducing build time brought by header files. Some of them optimize the header files themselves, by removing false code dependencies~\cite{yu2003removing}, modifying unnecessary include directives~\cite{spinellis2009optimizing, reisch2022automatic}, or replacing include directives with forward declarations~\cite{IWYU}. Others focused on optimizing the build processes through cache~\cite{koehler1995caching, koehler1997ccc}, precompilation~\cite{yu2005reducing} or detection of redundant compilation~\cite{dietrich2017chash}.
Besides, McIntosh et al.~\cite{mcintosh2016identifying} proposed an approach to identify header file hotspots, which undergo frequent modifications and trigger long-time rebuild processes. It helps developers identify and optimize header file hotspots, leading to reduced build times and increased productivity.

The above works have recognized the importance of header file optimization. Our work points out a new problem of God Header Files and gives a novel refactoring solution. We hope this will garner the interest of researchers and inspire the development of more effective approaches in the future.

\subsection{God Class Refactoring}

God Class is one of the most concerned code smells, referring to those complex classes that centralize the intelligence of the system~\cite{lanza2007object}. 
While some researchers question the necessity of refactoring God Classes~\cite{olbrich2010all, perez2014analyzing}, numerous studies have tackled this issue through semi-automatic~\cite{anquetil2019decomposing} or automatic approaches, primarily focusing on cohesion and coupling analysis. These studies utilize techniques such as static analysis, metric-based methods, and even deep learning approaches.
Bavota et al.~\cite{bavota2010playing, bavota2011identifying, bavota2014automating} proposed a series of static analysis based approaches that extract the God class by calculating the cohesion between two methods in the class. Akash et al.~\cite{akash2019approach} proposed a metric-based approach, applying Latent Dirichlet Allocation (LDA)~\cite{blei2003latent} to represent the methods by a distribution of topics. With the development of deep learning techniques, the utilization of Graph Neural Network model-based approaches has commenced to demonstrate its efficacy in addressing this particular task. A recent approach by Akash et.al~\cite{akash2022exploring} exploited the usage of graph auto-encoder to learn a vector representation for each method in the class.

We learned from how these works extract code relationships and incorporated build dependencies that is crucial in our task. We also utilize a GNN based graph clustering algorithm for improved decomposition.


\subsection{Graph Clustering}

Graph clustering, also known as community detection, is a fundamental problem in network analysis and has been extensively studied across various domains. 
Traditional algorithms are characterized by their reliance on mathematical and structural properties of graphs, such as spectral clustering~\cite{ng2001spectral}, modularity-based clustering like Louvain~\cite{blondel2008fast}, hierarchical clustering~\cite{ward1963hierarchical}, etc. However, these algorithms often lack scalability and struggle to handle large and complex graphs effectively.
In recent years, many deep learning-based graph clustering methods have been proposed. Such methods typically learn low-dimensional representations of graph nodes through Graph Neural Networks and then use them for clustering.
Some notable methods, such as GraphSAGE~\cite{hamilton2017inductive}, Graph Convolutional Networks (GCNs)~\cite{kipf2016semi}, and Variational Graph Autoencoders (VAGE)~\cite{kipf2016variational}, etc., outperform traditional clustering methods.

Despite these advancements, challenges remain due to the heterogeneity of real-world complex graph data, leading to the development of \textbf{multi-view graph clustering}. 
These works combine information from various sources or feature subsets.
By leveraging the consistency and complementary of different views, they exhibit superior effectiveness and generalization compared to single-view clustering.
These approaches encompass a wide spectrum, including multiple kernel clustering (MKC)~\cite{gonen2014localized, zhou2019multiple}, subspace clustering~\cite{brbic2018multi, li2019reciprocal}, NMF-based (non-negative matrix factorization) multi-view clustering~\cite{chen2020multi}, ensemble-based multi-view clustering~\cite{tao2019marginalized}, etc. Notable works such as O2MAC~\cite{fan2020one2multi}, MvAGC~\cite{lin2021graph}, MCGC~\cite{pan2021multi} and DuaLGR~\cite{ling2023dual} have achieved excellent performance. 
In our work, we employ DuaLGR due to its effectiveness with non-homophilous edges, dynamic weight assignment to address inconsistencies across projects, and scalability to incorporate additional relationships.

\section{Conclusion}
\label{sec:conclusion}

In this paper, we make aware of the problem of God Header File and propose an approach to automatically decompose God Header Files in C projects. 
We evaluate our approach on both synthetic God Header Files and real-world God Header Files. The results reveal that, in comparison to existing methods, our approach attains 11.5\% higher accuracy and demonstrates more consistent performance across different projects. When applied to real-world God Header Files, our decomposition results exhibit better architecture, evidenced by higher modularity and acyclic dependencies, and achieve a 15\% to 60\% reduction in recompilation.
In future work, we will explore how to recommend the optimal number of decomposed files, generate the appropriate file name, and refactor all the related files automatically, facilitating code comprehension and maintenance in C projects.

\section*{Acknowledgment}
This work is supported by the National Key Research and Development
Program of China under Grant No.2023YFB4503803.


\balance
\bibliographystyle{IEEEtran}
\bibliography{main}

\end{document}